# Agglomeration of Charged Nanopowders in Suspensions

J.H. Werth[1], M. Linsenbühler[2], S.M. Dammer[1], Z. Farkas[1], H. Hinrichsen[1,3], K.-E. Wirth[2], and D.E. Wolf[1]

[1] Institut für Physik, Gerhard-Mercator-Universität Duisburg, 47048 Duisburg, Germany
[2] Mechanische Verfahrenstechnik, Universität Erlangen-Nürnberg, 91058 Erlangen, Germany
[3] Fachbereich Physik, Bergische Universität Wuppertal, Gaußstrasse 20, 42097 Wuppertal, Germany

**Abstract**

The aim of this work is to understand agglomeration of charged powders suspended in nonpolar fluids. The concerted influence of electromagnetic, hydrodynamic and van der Waals forces as well as Brownian motion leads to a complex agglomeration behaviour which depends on several parameters, e.g., the ratios of electric charges, particle sizes, temperature and concentrations of the particles. Both experimental and theoretical considerations are presented.

## 1. Introduction

Granular powders are essential for the fabrication of pharmaceutical products. An increasingly important application is the design of inhalable drugs, where particles on the micrometer scale are deposited in the alveoli of the lung. A major problem in this context is the clumping of small particles due to van der Waals forces, leading to large agglomerates during the dispersion process. A possible method to avoid clumping would be to coat the particles by smaller nanoparticles, as sketched in Fig. 1. The small particles act as spacers between the larger particles, thereby reducing the mutual van der Waals forces.

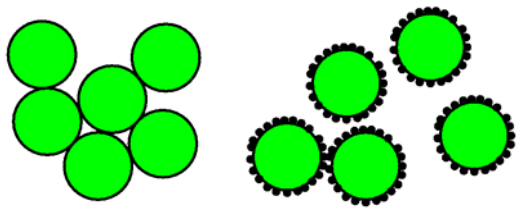

**Fig. 1:** *Reduction of van der Waals forces by coating larger particles with small nanoparticles.*

Optimal results are obtained if the surface density of the coating particles is uniform and neither too small nor too large. To this end one has to avoid that the nanoparticles themselves aggregate before they are deposited. A promising technological approach, on which we will focus in the present work, is to charge both particle fractions oppositely.

While the electrical charges stabilize each fraction against agglomeration, they support the coating process as the bigger particles attract the nanoparticles.

This coating strategy requires an insulating, nonpolar liquid. In the experiments reported here we used liquid nitrogen. The particles are charged by two different methods, namely, triboelectric as well as contact charging. Using Aerosil powders as a model substance, we find that the electrostatically supported coating process is feasible and improves the flowability of the powder considerably.

The deposited layer, however, is not as simple as Fig. 1 suggests. Scanning electron microscopy (SEM) pictures show that the large particles are coated by *agglomerates* instead of single nanoparticles. In order to understand the complex coating process in more detail and to optimize the experimental parameters we perform extensive molecular dynamics simulations. Moreover we present physical arguments concerning the stability of a charged suspension and the dynamics of the aggregation process.

## 2. Charging process and experimental setup

The experimental investigations were done at the Department of Powder Technology at the University of Erlangen-Nuremberg. Several devices to charge powders electrostatically in liquid nitrogen were built. Two different mechanisms are used to charge the particles, each with its advantages and disadvantages: triboelectric and contact charging.

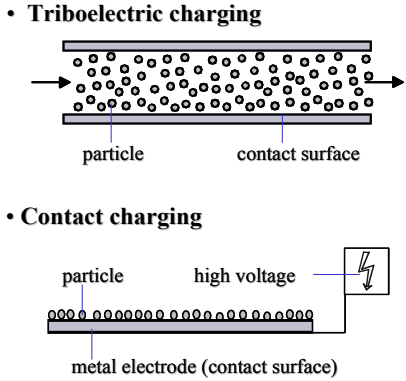

**Fig. 2:** *Charging concepts used in the experiments.*

**Triboelectric charging:** If two electrically neutral insulators with different electrostatic properties come into contact (cf. upper panel of Fig. 2), they transfer charges from surface to surface in order to approach thermodynamic equilibrium, i.e., to reduce chemical potential differences. If the surfaces are separated quickly e.g. by sliding or rubbing, an excess charge will remain on the surfaces. In the experiment the particles are charged by collisions with a rotating stirrer, which is also used to disperse the powder. Thus triboelectric charging in liquid nitrogen has the advantage that the charging process and the dispersion of the powder can be carried out simultaneously by using a high shear agitating system.

**Contact charging:** The principle of contact charging is sketched in the lower panel of Fig. 2. In our experimental set-up the suspension is pumped through a metal electrode in form of a grid connected to a high-voltage supply. Because of the potential difference electric charges are transferred from the electrode to the particles. To optimise this method we developed a U-shaped charging device, where the suspension moves periodically back and forth in order to pass the electrodes several times.

**Comparison of the methods:** Previous experiments by Huber [1] showed that triboelectric charging is more efficient in the sense that higher surface charges per particle can be reached. Moreover, this technique needs less experimental effort than contact charging. Regarding the coating problem triboelectric charging is particularly advantageous, if the big and the small particles consist of different materials so that they are charged oppositely, when the suspension containing both of them is stirred. However, triboelectric charging has the disadvantage that the polarity depends on the used materials and cannot be controlled in the experiment.

**Surface charge densities in the experiment:** As an example, where both particle fractions are oppositely charged during the dispersion process, we studied the deposition of Aerosil R972 (Degussa, $d_p$ = 0,012 µm) on Lactose ($d_p$ = 5 µm). The particles were charged in liquid nitrogen with a high speed dispersion system (Ultra-Turrax-Disperser IKA T50). The specific charge of the two different powders as function of the charging time is shown in Fig. 3: Lactose accumulates a positive charge density (approximately 150 nC/m²) on the surface of its particles, while Aerosil R972 was charged negatively (approximately -280 nC/m²).

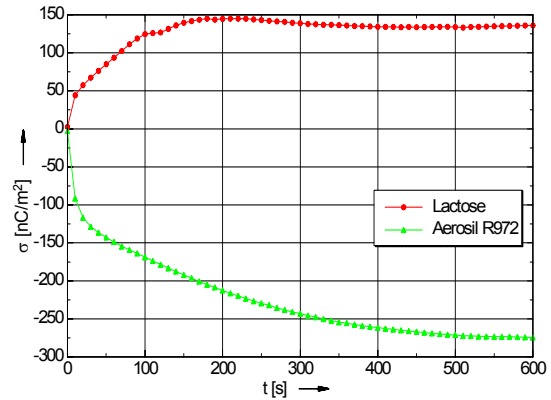

**Fig. 3:** *Surface charge density as a function of the charging time [2]*

## 3. Simulation model

In order to study the stabilizing effect of the Coulomb repulsion theoretically let us consider an idealized suspension of monopolarly charged spherical particles. The particles in the suspension are subjected to various forces, which are of electrical (van der Waals and Coulomb forces) and of hydrodynamic origin (Brownian motion, Stokes friction and long-range hydrodynamic forces). In our molecular dynamics simulations, all these forces are taken into account.

**Van der Waals forces** are caused by fluctuating electric dipoles and decay asymptotically as $r^{-7}$, where $r$ is the distance between the particle centers. On short distances they diverge like $d^{-2} = (r-2a)^{-2}$, where $a$ is the particle radius and $d$ is the distance between the surfaces. For particles smaller than a few micrometers the van der Waals force dominates all other forces on short distances. Therefore, we implemented this interaction as a purely adhesive force, i.e., as long as the particles are separated their van der Waals interaction is neglected, but if they collide they stick together irreversibly.

**Electrostatic forces:** Since the liquid is non-polar and free of countercharges, the $r^{-2}$-asymptotics of the Coulomb forces is not screened in the present case. Therefore, the long-range part is implemented without further approximations. On short distances the situation is more complex since the charges are

2/7

located on the surfaces of the particles and may come into contact when two particles approach each other so that the Coulomb force diverges. But as long as both particles carry charges of the same polarity, the approaching particles will rotate in such a way that the distance between the charges is maximized. Thus in the case of monopolarly charged suspensions the Coulomb force cannot diverge, justifying the approximation to locate the charges in the centers of the particles.

**Hydrodynamic forces:** In order to reduce the computational effort, we do not simulate the liquid explicitly. Instead, we consider effective forces between the particles using the Stokeslet approximation [3, 4]. This approximation is valid for low particle concentrations and works as follows: First, approximate velocities $\vec{v}_{0,i}$ are determined for each particle $i$ by balancing Stokes friction with the Coulomb force exerted by the other charged particles,

$$\vec{F}_i^{Coulomb} = -\vec{F}_i^{Stokes} = 6\pi\eta a \vec{v}_{0,i}. \quad (1)$$

Each moving particle generates a velocity field in the surrounding liquid. These fields can be linearly superposed in the limit of low Reynolds number. This is justified since the Reynolds number for liquid nitrogen on the typical scale of the suspended particles is $Re \approx 4 \cdot 10^{-5}$. For a spherical particle with radius $a$ located at the origin and moving at velocity $\vec{v}_0$, the corresponding approximated velocity field $\vec{u}(\vec{r})$ in the liquid at position $\vec{r}$ is given by

$$\vec{u}(\vec{r}) = \frac{3}{4}a\left(\frac{\vec{v}_0}{r} + \frac{(\vec{v}_0 \cdot \vec{r})\vec{r}}{r^3}\right). \quad (2)$$

In the limit of linear hydrodynamics these velocity fields generated by the various particles can be superposed and result in a fluid velocity at the position of particle $i$ given by

$$\vec{U}_i = \sum_{j \neq i} \vec{u}_j(\vec{r}_i - \vec{r}_j), \quad (3)$$

which for dilute suspensions is a small correction to the velocity $\vec{v}_{0,i}$ in the local rest frame of the liquid. Hence the velocity of particle $i$ in the laboratory frame in Stokeslet approximation is given by

$$\vec{v}_i = \vec{v}_{0,i} + \vec{U}_i. \quad (4)$$

Thus, given the particle positions at time $t$, the procedure to calculate the new particle positions at time $t+\Delta t$ reads as follows:

(i) Calculate all Coulomb forces acting on each particle

$$\vec{F}_i^{Coulomb} = \sum_{j \neq i} \frac{q_j}{4\pi\varepsilon\varepsilon_0(\vec{r}_i - \vec{r}_j)},$$

where $q_j = 4\pi a^2 \sigma$.

(ii) Compute $\vec{v}_{0,i}$ for each particle $i$ according to (1).

(iii) Compute $\vec{v}_i$ for each particle $i$ according to (2)-(4).

(iv) The positional shift $\vec{r}_i(t + \Delta t) - \vec{r}_i(t)$ of each particle is given by $\vec{v}_i \cdot \Delta t + \vec{\xi}_i$, where $\vec{\xi}_i$ is the shift due to Brownian motion determined from a Gaussian distribution of random numbers with the correlations

$$\langle \xi_i(t)\xi_j(t')\rangle = 2D\Delta t \delta_{ij}\delta(t - t').$$

The diffusion constant $D$ is connected with temperature, viscosity and particle radius by the Einstein relation $D = k_B T / 6\pi\eta a$, where $k_B$ is the Boltzmann factor.

Note that in our approximation the particle motion is overdamped, i.e. the velocities of particles are proportional to forces. This is justified by a comparison of the velocity relaxation time, $t_{relax} = m/6\pi\eta a$, and the typical time a particle needs to diffuse a distance equal to its own diameter, $t_{diff} = 2a^2/D = 18\pi\eta a^3/k_B T$. Inserting the particle mass $m = 4\pi a^3 \rho_m / 3$, where $\rho_m$ is the mass density, we find that for typical experimental conditions $t_{relax}$ is always much smaller than $t_{diff}$, even if one considers particles as small as $a = 1$ nm.

## 4. Charge stabilization

Let us now address the question under which conditions a suspension of equally charged particles is stable. Starting the simulation with homogeneously distributed particles one observes a quick formation of aggregates. After a certain time the flakes reach a characteristic size from where on they grow much slower (see Fig. 4) so that the suspension is practically stable. We argue that the threshold, from where on a monopolarly charged powder is stable, can be estimated by a simple comparison of energies. If the energy needed to bring two oppositely charged agglomerates in contact (effective Coulomb-barrier) is smaller than the thermal energy, the particles will aggregate (unstable phase), otherwise they will remain separated (stable phase). Because of the vicinity of the other charged particles, this energy depends on the density of agglomerates.

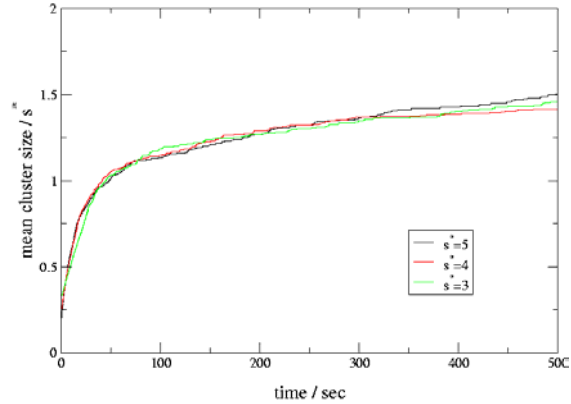

**Fig. 4:** *Mean cluster size divided by the characteristic cluster size s\* depending on time.*

For spherical particles the effective Coulomb barrier can be expressed as [5]

$$E_C^{eff} = \frac{1}{4\pi\varepsilon\varepsilon_0}\frac{q^2}{2a}\left(1 - \frac{2a}{l}\right)^2, \quad (5)$$

where $\varepsilon$ is the dielectric constant of the surrounding liquid (for liquid Nitrogen $\varepsilon \approx 1.45$), $\varepsilon_0$ is the electric constant of vacuum, $l$ is the typical distance between particles, $a$ is their radius, and $q = 4\pi a^2 \sigma$ is the charge of the particles which is assumed to be proportional to the surface charge density $\sigma$. Fig. 5 shows schematically the energy relations and explains the condition of clustering according to this energy argument. The characteristic temperature $T^*$, where the system is at the borderline between the stable and the unstable regime, is given by[1]

$$3k_B T^* = E_C^{eff}. \quad (6)$$

The relevant parameters are the temperature of the liquid, the surface charge density, the radius, and the density of the particles. For the density we use the volume fraction $\rho$, which, assuming an average BCC lattice, is given by $\rho = \sqrt{3}\pi(a/l)^3$. This relation also holds for other regular lattices with a different prefactor. Note that in the energy argument the viscosity does not enter since in the overdamped limit it only rescales time. Hence the characteristic temperature is

$$T^*(a,\sigma,\rho) = c_1^{-1}\sigma^2 a^3 (1-c_2 \rho^{1/3})^2 \quad (7)$$

with $c_1 = 3k_B \varepsilon\varepsilon_0/(2\pi)$ and $c_2 = 2(\sqrt{3}\pi)^{-1/3}$. Equation (7) predicts a characteristic particle radius (see Fig. 6)

$$a^*(T,\sigma,\rho) = c_1^{1/3} T^{1/3} \sigma^{-2/3} (1-c_2 \rho^{1/3})^2 \quad (8)$$

which means that particles of smaller radius tend to aggregate, while a suspension of larger particles is stable at a given temperature $T$.

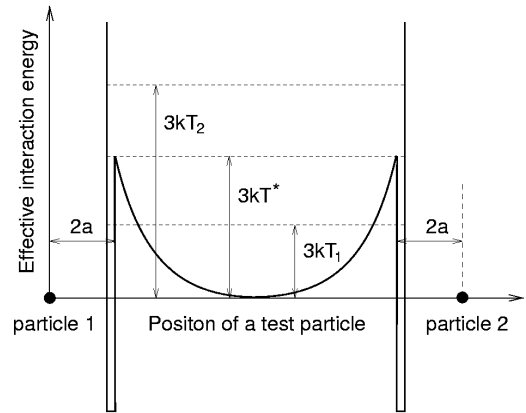

**Fig. 5:** *The effective interaction energy between a test particle and the other particles (here denoted as particle 1 and 2) as a function of the test particle position. The distance between the particles is 2l. T\* is the characteristic temperature, where the average thermal energy is large enough to overcome the repelling electrostatic force. Below this temperature ($T_1$) the particles can only occasionally overcome the potential barrier so that most of them remain separated, while for higher temperatures ($T_2$) they aggregate. The potential wells at distance 2a from the particle centers represent the van der Waals attraction and are deep enough to ensure that the aggregation is irreversible.*

---

[1] Note that rotational degrees of freedom need not to be taken into account and were actually not simulated. The left hand side of Eq. (6) is the thermal energy of the relative motion of *two* particles.

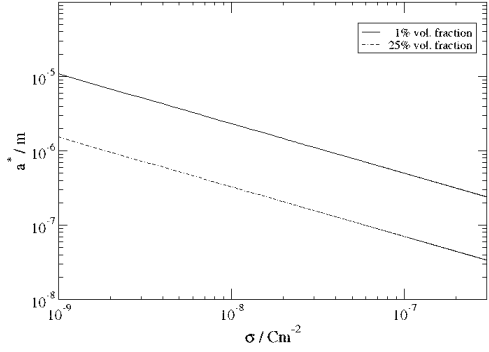

**Fig. 6:** *Critical particle radius as function of the specific charge and the volume fraction.*

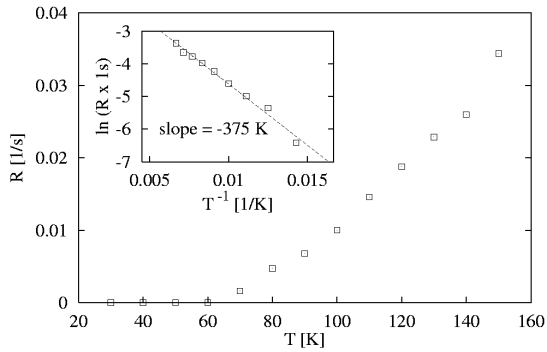

**Fig. 7:** *The clustering rate R of unipolarly charged, equally sized particles as a function of temperature T. Typical experimental values are chosen as parameters: $a = 0.5\ \mu m$, $\eta = 10^{-4}\ Ns/m^2$, $\sigma = 2.7 \times 10^{-7}\ C/m^2$, $\varepsilon = 1$, $\rho = 0.0067$ [6]. Inset: Arrhenius plot of the same data. The negative slope, i.e. the activation energy, has to be identified with $3T^*$ according to Eq. (6). The simulation result $T^* \approx 125K$ is of the same order of magnitude as the value calculated from (7), $T^* = 96.4K$.*

Regarding the aggregation dynamics, the competition between Coulomb energy and thermal energy influences the collision rate of the particles. Here it turns out that the clustering rate in the aggregation regime is proportional to

$$\exp(-E_C^{eff}/3k_BT).$$

The derivation follows the same arguments as in the case of granular gases [5], where it had been shown that the collision rate between charged particles is exponentially suppressed. In that case the exponent is given by the ratio of the Coulomb barrier and the average kinetic energy of the grains, the so-called granular temperature. Thus the characteristic temperature $T^*$ in Eq. (7) marks the crossover between a regime, where the clustering rate is fast from a regime in which the clustering rate is slow. This is confirmed by our simulation results presented in Fig. 7 which shows the initial clustering rate in a dilute suspension of spherical particles. Within the simulation time no significant clustering was observed for small temperatures, confirming that the suspension was indeed stabilized by Coulomb repulsion.

For high temperatures clusters continue to grow, thereby becoming more and more charged. Hence, the Coulomb barrier increases, which corresponds to an increasing characteristic temperature $T^*$. When the typical Coulomb barrier $3k_BT^*$ becomes larger than the thermal energy $k_BT$, which happens when the clusters reach a characteristic size $s^*$, further aggregation is suppressed by Coulomb repulsion. Thus for high temperatures we expect initial power law growth of the cluster sizes which should turn over to much slower growth when the cluster size reaches $s^*$. The computer simulations support this prediction: In Fig. 4 we divided the mean clustersize by $s^*$ in order to demonstrate that all three systems cross over to the slow growth regime at this value.

### 5. Coating process

Let us finally discuss the experimental procedure to coat large particles by smaller ones. As described in Section 2 we use a mixture of Lactose and Aerosil. The process flow is shown in Fig. 8. By stirring the suspension the coarse and the fine fraction are oppositely charged. Because of the electrostatic attraction between the fine and the coarse particles the Lactose is coated by Aerosil. As stated earlier, the fine fraction, which is deposited on the surface of big particles, consists of agglomerates instead of single particles. After evaporating the liquid nitrogen the agglomerated dry powder is obtained.

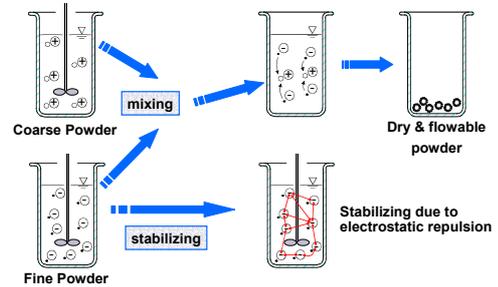

**Fig. 8:** *Process flow diagram*

After the dispersion and charging procedure we measured the agglomerate size of the nano-material. One possibility is to measure the sizes directly on the coated carrier particles (see Fig. 9). Alternatively, it is also possible to deposit them on a TEM-grid and to analyze the shape and the size of the deposited Aerosil agglomerates.

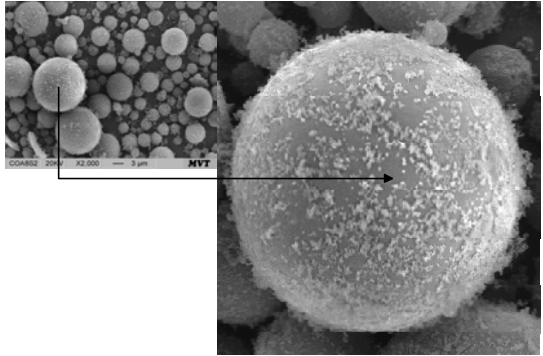

**Fig. 9:** *SEM-picture of lactose coated with Aerosil [1].*

The analysis of the Fig. 9 provides an estimated flake size of 150 nm. The used SEM is not able to deliver the required resolution for an exact analysis of the agglomerates. Therefore, the agglomerate size distribution was investigated using a TEM microscope (PHILIPS CM 30 T/STEM). The TEM-grids were prepared by cooling them down in liquid nitrogen, so that the temperature difference between the suspension and the grid was small. Then the TEM-grids were coated by dipping them directly into the suspension. Due to thermophoresis the particles were deposited on the grid. Two TEM-pictures of Aerosil R972 with different magnifications are shown in Fig. 10. The left picture shows a couple of clusters with different flake-sizes and different primary-particle sizes. Remarkably, some of the flakes seem to be bolder than others because of a different degree of sintering during their production in the flame reaction chamber.

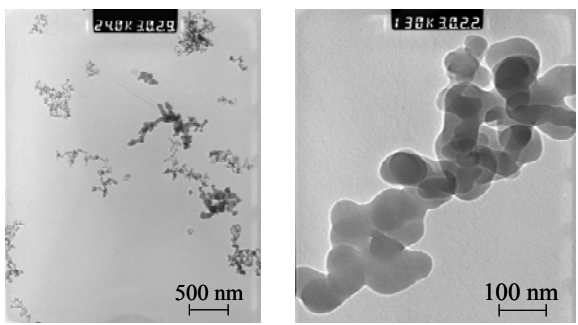

**Fig. 10:** *TEM-picture of Aerosil flakes*

The picture in Fig. 10 on the right shows a sintered flake of Aerosil particles. Analysing Aerosil R972 no primary particles were detected in the suspension. Thus the dispersion process was not able to destroy the flakes by shearing force or by electrostatic force.

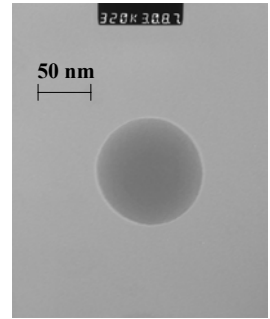

**Fig. 11**: *TEM-pictures of single Aerosil OX50 nano-particle.*

In order to investigate suspensions of isolated primary particles one has to resort to other substances. First results with Aerosil OX50 show that single primary particles can be detected in suspensions of this substance, as shown in Fig. 11. The aim of future investigations will be to use this substance in order to verify the theoretical predictions, especially the characteristic radius of the aggregates produced by the Brownian motion.

## 6. OUTLOOK

We have reported experimental and theoretical studies on the behavior of electrically charged nanoparticles suspended in liquid nitrogen. After dispersing and charging the particles in an appropriate facility we have investigated how they agglomerate. Liquid nitrogen was used as dispersion medium for several reasons [1], in particular because there is no hydration of the particles.

As the experiments have shown (see Fig. 10) presintered flakes cannot be broken up in the dispersion process. Hence, in contrast to the simulation, the suspension contains initially a distribution of flakes with different sizes and probably also different charges. We postulate that these flakes further aggregate until they reach a typical size given by our theoretical argument. Thus we believe that our main conclusion, namely, the existence of characteristic flake size for given charge and temperature, remains valid even for partially agglomerated nanopowders, provided that the characteristic flake size is larger than the primary aggregates of the freshly dispersed suspension. In the future we intend to perform simulations with partially aggregated initial state. Moreover we intend to carry out experiments with suspensions containing isolated primary particles in order to comply with the simulation results.

Presently it is not yet clear whether the flakes deposited on the surface of the Lactose particles in Fig. 9 correspond to the presintered flakes or whether further agglomeration happened before the deposition. To clarify this question is important with respect to the coating process.

The simulation results presented in this paper were carried out with moderate computational effort. In order to study statistical properties like the cluster size distribution quantitatively, large scale computations on parallel computers are needed. However, a parallelization of the program is a nontrivial task because of the long-range interactions. One possible solution was presented in [7], where a hypersystolic algorithm was used.

In summary we have shown that a characteristic and stable cluster size exists which is caused by the interplay of Coulomb forces and Brownian motion. The characteristic cluster size depends on the suspension temperature and the solid concentration.

## 7. List of Symbols

| | |
|---|---|
| a | particle radius (m) |
| D | diffusion constant (J m/N s) |
| ε | dielectric constant (F/m) |
| E | energy (J) |
| η | viscosity (N s/m²) |
| $k_B$ | Boltzmann constant (J/K) |
| l | typical distance between neighbouring particles (m) |
| m | particle mass (kg) |
| q | charge of a particle (C) |
| ρ | number density of particles in suspension ($m^{-3}$) |
| $\rho_m$ | mass density (kg $m^{-3}$) |
| r | distance from the center of a particle (m) |
| s | characteristic cluster size |
| $\vec{u}(\vec{r})$ | velocity field generated by a particle (m/s) |
| $\vec{U}_i$ | total velocity field felt by particle $i$ (m/s) |
| $v_i$ | velocity of particle $i$ (m/s) |
| σ | surface charge density (C/m²) |
| t | time (s) |
| T | temperature (K) |
| ξ | shift due to Brownian motion (m) |

## 8. Acknowledgement

Financial support of the theoretical and experimental investigations by the Deutsche Forschungsgemeinschaft (DFG), projects WI 972/14-1 and HI 744/2-1, is gratefully acknowledged.